\documentclass[%
 reprint,
 amsmath,amssymb,
 aps,
 superscriptaddress,
]{revtex4-1}

\usepackage{graphicx}% Include figure files
\usepackage{dcolumn}% Align table columns on decimal point
\usepackage{bm}% bold math
\begin{document}

\preprint{APS/123-QED}

\title{\textit{LISA} Sources in Milky Way Globular Clusters}% Force line breaks with \\

\author{Kyle Kremer}
 \email{kremer@u.northwestern.edu}
\affiliation{
 Department of Physics \& Astronomy, Northwestern University, Evanston, IL 60202, USA\\
 }
\affiliation{
 Center for Interdisciplinary Exploration \& Research in Astrophysics (CIERA), Evanston, IL 60202, USA\\
}

\author{Sourav Chatterjee}%
 \email{chatterjee.sourav2010@gmail.com}
\affiliation{
 Department of Physics \& Astronomy, Northwestern University, Evanston, IL 60202, USA\\
 }
\affiliation{
 Center for Interdisciplinary Exploration \& Research in Astrophysics (CIERA), Evanston, IL 60202, USA\\
}
\affiliation{Tata Institute of Fundamental Research, Homi Bhabha Road, Mumbai 400005, India}

\author{Katelyn Breivik}
\affiliation{
 Department of Physics \& Astronomy, Northwestern University, Evanston, IL 60202, USA\\
 }
\affiliation{
 Center for Interdisciplinary Exploration \& Research in Astrophysics (CIERA), Evanston, IL 60202, USA\\
}

\author{Carl L. Rodriguez}
\affiliation{%
 MIT-Kavli Institute for Astrophysics and Space Research, Cambridge, MA 02139, USA \\
}%

\author{Shane L. Larson}
\affiliation{
 Department of Physics \& Astronomy, Northwestern University, Evanston, IL 60202, USA\\
 }
\affiliation{
 Center for Interdisciplinary Exploration \& Research in Astrophysics (CIERA), Evanston, IL 60202, USA\\
}

\author{Frederic A. Rasio}
\affiliation{
 Department of Physics \& Astronomy, Northwestern University, Evanston, IL 60202, USA\\
 }
\affiliation{
 Center for Interdisciplinary Exploration \& Research in Astrophysics (CIERA), Evanston, IL 60202, USA\\
}

\date{\today}% It is always \today, today,

\begin{abstract}
We explore the formation of double-compact-object binaries in Milky Way (MW) globular clusters (GCs) that may be detectable by the Laser Interferometer Space Antenna (\textit{LISA}). We use a set of 137 fully evolved GC models that, overall, effectively match the properties of the observed GCs in the MW. We estimate that, in total, the MW GCs contain $\sim 21$ sources that will be detectable by \textit{LISA}. These detectable sources contain all combinations of black hole (BH), neutron star, and white dwarf components. We predict $\sim 7$ of these sources will be BH--BH binaries. Furthermore, we show that some of these BH--BH binaries can have signal-to-noise ratios large enough to be detectable at the distance of the Andromeda galaxy or even the Virgo cluster.
\end{abstract}
%\begin{description}
%\end{description}
%\end{abstract}

\maketitle

%\tableofcontents

\section{\label{sec:intro}Introduction}
Over the past several decades, an increasing number of binary systems containing compact objects, including white dwarfs (WDs) \citep[e.g.,][]{Moehler2008}, neutron stars (NSs) \citep[e.g.,][]{Giacconi1974, Clark1975b, Camillo2000}, and black holes (BHs) \citep[e.g.,][]{Strader2012, Chomiuk2013,Giesers2018}, have been observed and studied in globular clusters (GCs). Although present-day GCs typically contain a small fraction of the total stellar mass of their host galaxies, dynamical formation channels unique to GCs produce an overabundance of close compact-object binaries per mass, relative to the Galactic field \citep[e.g.,][]{Katz1975, Clark1975,Hut1992}. The dynamical formation of these systems has been studied extensively using various computational methods \citep[e.g.,][]{Bailyn1995, Mackey2008, Chatterjee2013, Morscher2013,Rodriguez2016b, Antonini2016,Kremer2017a}.

Recent analyses have shown that GCs may be a dominant formation channel for the BH--BH binaries observed by the Laser Interferometer Gravitational-Wave Observatory (LIGO) \citep[e.g.,][]{Banerjee2010,Ziosi2014,Rodriguez2015,Rodriguez2016a,Chatterjee2017a,Chatterjee2017b}. Before entering the high-frequency range of ground-based gravitational wave detectors like LIGO, these BH--BHs will be detectable as low-frequency
sources by the Laser Interferometer Space Antenna (\textit{LISA}) \citep{Amaro2013,Amaro2017}. 

In addition to BH--BH binaries, BH--NS and NS--NS binaries will also be detectable by \textit{LISA}. Furthermore, unlike LIGO, \textit{LISA} will also be able to observe compact binaries with WD components, including WD--WD binaries, which make up the largest fraction of close compact-object binaries in the Milky Way (MW) \citep[e.g.,][]{Marsh1995}.

Several previous studies have noted that GCs may harbor populations of compact binaries detectable by \textit{LISA} \citep[e.g.,][]{Benacquista2001, Willems2007}. Additional studies have also shown that young open clusters may also contribute to the number of dynamically formed \textit{LISA} sources in the MW \citep[e.g.,][]{Banerjee2017,Banerjee2018}. Furthermore, it has been shown that the dynamical formation channels unique to clusters may produce populations of \textit{LISA} sources with orbital features (e.g., eccentricity) that are very different from \textit{LISA} sources formed in the Galactic field \citep[e.g.,][]{Willems2007,Breivik2016}. 

In this analysis, we explore the formation of \textit{LISA} sources within MW GCs. Using our cluster Monte Carlo code, \texttt{CMC}, we utilize a set of 137 fully evolved GC models with present-day properties similar to those of the observed GCs in the MW. We identify potential \textit{LISA} sources created within these models, including those retained within their host GCs at present and those ejected into the Galactic halo. 

In Section \ref{sec:method}, we describe our computational method to model GCs. %and discuss how we determine whether a particular binary would be observable by \textit{LISA}.
In Section \ref{sec:retained}, we show all compact object binaries retained in their host clusters at late times and estimate the total number of \textit{LISA} sources predicted to be found in MW GCs, as well as GCs in Andromeda and the Virgo cluster. %In Section \ref{sec:ejected}, we examine the contribution of ejected compact object binaries to the \textit{LISA} population.
We conclude in Section \ref{sec:conclusion}.

\section{Globular cluster models}
\label{sec:method}

%\subsection{Globular cluster models}
%\label{sec:models}
We compute our GC models using \texttt{CMC}, Northwestern's cluster Monte Carlo code \citep[e.g.,][]{Joshi2000,Joshi2001,Fregeau2003, Fregeau2007, Chatterjee2010,Chatterjee2013, Pattabiraman2013,Umbreit2012,Morscher2013,Morscher2015,Rodriguez2016b}. \texttt{CMC} includes all physics relevant for studying the formation and evolution of compact-object binaries in dense star clusters, such as two-body relaxation \citep{Henon1971a,Henon1971b}, strong binary-mediated scattering \citep{Fregeau2004}, single and binary star evolution (implemented using the single and binary star evolution software packages, \texttt{SSE} and \texttt{BSE} \citep{Hurley2000,Hurley2002,Chatterjee2010,Kiel2009}), and Galactic tides \citep{Chatterjee2010}. Note that \texttt{SSE} and \texttt{BSE} adopt several simplifications, especially for dynamically modified stars (such as stellar collision products). Nonetheless, \texttt{SSE} and \texttt{BSE} are widely accepted as state of the art and used in most dynamics codes. A more realistic approach may involve individually evolving stars using, for example, \texttt{MESA}. However, this is beyond the scope of current simulations both due to stability issues and computational cost.

Here we use the set of 137 GC models listed in \citep{Kremer2017b}. For these models, a number of initial cluster parameters are varied, including particle number, virial radius, King concentration parameter, and binary fraction, among others.
The full set of models and initial conditions are listed in the Appendix of \citep{Kremer2017b}.

Each model features a set of snapshots in time spaced 10--100 Myr apart that list the orbital parameters of all binaries in the GC at each point in time. Here, we look at all snapshots between 11--12 Gyr (the same age range considered in \citep{Breivik2016}), to reflect the observed age spread of MW GCs. Each of these snapshots is considered to be an equally valid representation of a GC at late times, similar to the old GCs observed in the MW. Note that the tail of the MW GC age distribution may extend down to $\sim 9$ Gyr \citep[e.g.,][]{Lee1994,Milone2014}. However, we find that extending our age range down to 9 Gyr has no significant effect on the number of \textit{LISA} sources.

For each model, we count the total number of compact object binaries found in each time snapshot and determine how many of these binaries may be found within the \textit{LISA} sensitivity range. 

%In order to make predictions of the total number of \textit{LISA} sources in the MW, we use the weighting scheme implemented in \citep{Kremer2017b}, where each model receives a weight based on how well it matches observed MW clusters in the $r_c-r_h$ plane (see Section 5.2 of \citep{Kremer2017b} for more detail). After applying the weighting scheme, we re-scale the weighted models to match the total mass of the observed MW GCs.

In order to estimate the total number of \textit{LISA} sources in the MW, we rescale these numbers based upon the total number of snapshots found in each model in the range 11--12 Gyr. Additionally, we implement the weighting scheme of \citep{Kremer2017b}, where each model is weighted based on how well it matches observed MW clusters in the $M_{\rm{tot}}-r_c/r_{hl}$ plane, where $M_{\rm{tot}}$ is the total mass of each cluster and $r_c$ and $r_{hl}$ are the observed core and half-light radius, respectively (see Section 5.2 of \citep{Kremer2017b} for more detail). After applying the weighting scheme, we rescale the weighted models to match the total mass of the observed MW GCs. As in \citep{Kremer2017b}, we also implemented an alternative weighting scheme where the models are instead weighted in the $r_c-r_{hl}$ plane, and find that our results do not change significantly.

%Initial galactocentric distances for our set of models vary from $2-20$ kpc (see Table 3 of \citep{Kremer2017b}).  However, our GC models are not tidally filling, so tidal truncation is not a dominant driving factor, particularly for compact objects which preferentially reside in their host cluster’s center due to mass segregation (see \citep{Chatterjee2013} for further discussion of tidal treatment in our models). Therefore, present-day heliocentric distances for each GC model are randomly drawn from the observed heliocentric distances of MW clusters (taken from \citep{Harris1996}) and the $S/N$ for all binaries in each model is calculated at the randomly selected distance. This allows us to sample \textit{LISA} signal-to-noise ratios from a distance distribution representative of observed MW clusters. We draw 10 random heliocentric distances for each model, and then re-scale our final results down by a factor of 10.

Initial Galactocentric distances for our set of models vary from 2 to 20 kpc (see Table 3 of \citep{Kremer2017b}). However, none of our GC models are initially tidally filling (a well-known outcome of choosing initial properties, such as the virial radius, guided by the observed young super star clusters; see \citep{Chatterjee2013}). Although tidal truncation can become important at late times, particularly for clusters near the Galactic center, compact objects preferentially reside in their host cluster’s centers due to mass segregation; thus, the dynamics for them is not significantly affected by the choice of Galactocentric distance.  Thus, we treat each model as a representative dynamical factory for the creation of compact-object binaries and ignore the particular choice of Galactocentric distance for each model.

However, to calculate the signal-to-noise ratio ($S/N$) expected for \textit{LISA}, we need to assign \textit{heliocentric} distances for the sources. To be realistic in these distance assignments, we directly sample from the present-day heliocentric distances of the MW GCs \citep{Harris1996} and assign all sources created in particular models this heliocentric distance. To account for statistical fluctuations, we draw ten randomly sampled distances for each model from the heliocentric-distance distribution of the MW GCs, and then rescale our final results down by a factor of 10.

During the evolution of a GC, many compact-object binaries will be ejected from the cluster as the result of dynamical encounters, supernova (SN) explosions, and tidal stripping \citep[e.g.,][]{Kremer2017b}. Depending upon the time of ejection and the orbital parameters of these binaries at the time of ejection, such binaries may be observed as \textit{LISA} sources at the present day as members of the Galactic halo. In order to estimate the number of such systems, we evolve all ejected binaries forward in time using \texttt{BSE} to determine the properties at the present day.

%\subsection{Calculation of signal-to-noise ratio}
\begin{figure*}
\begin{center}
\includegraphics[width=0.85\textwidth]{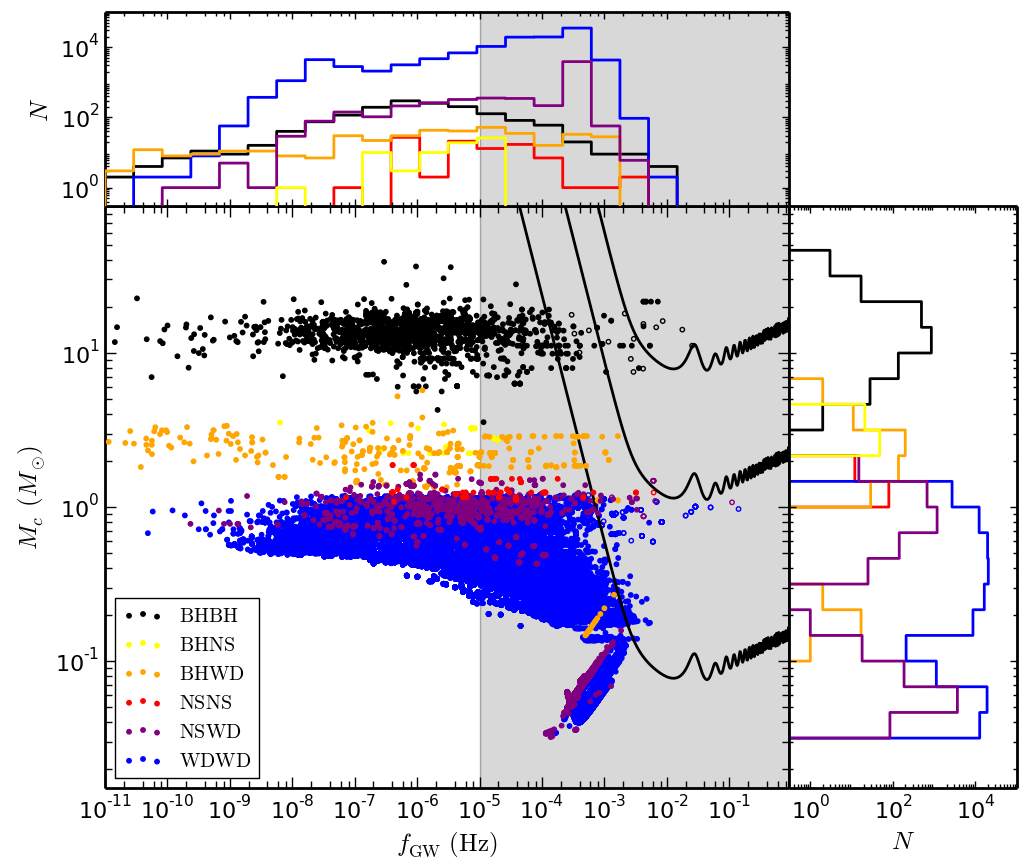}
\caption{\label{fig:retained} All compact-object binaries found in our set of  models at late times (defined here as $11 \, \rm{Gyr} < \it{t} < \rm{12} \, \rm{Gyr}$). Black systems mark BH--BH binaries, yellow BH--NS binaries, orange BH--WD binaries, red NS--NS binaries, purple NS--WD binaries, and blue WD--WD binaries. The background gray region marks the \textit{LISA} sensitivity range. The three solid black curves mark the boundary for detecting systems with $S/N \geq 2$ at distances of $d=9$ kpc (bottom; median heliocentric distance of MW GCs), $d=800$ kpc (middle; distance to Andromeda), and $d=1.6\times10^4$ kpc (top; distance to the Virgo cluster). The height of each point above each respective curve scales as $(S/N)^{3/5}$. Note that this figure shows results for \textit{all snapshots} in the range 11 Gyr $< t<$ 12 Gyr for our models, meaning this figure overcounts the number of binaries by a factor equal to the number of snapshots ($\sim 10$) between 11 and 12 Gyr for each model.}
\end{center}
\end{figure*}

\section{Retained Systems}
\label{sec:retained}

To compute the \textit{LISA} sensitivity for the binaries considered in this analysis, we model gravitational waves (GWs) at the leading quadrupole order, as detailed in the Appendix. We adopt $T_{\rm{obs}}=4$ years as the \textit{LISA} observation time.

Because many compact-object binaries in GCs will have high eccentricities induced by dynamical encounters, it is useful to consider the frequency of GWs emitted at higher harmonics than the orbital frequency. The frequency of maximum GW power emission from an eccentric binary is given by \citep{Wen2003} as

\begin{equation}
f_{\rm{GW}} = \frac{\sqrt{G(M_1 + M_2}}{\pi}\frac{(1+e)^{1.1954}}{[a\,(1-e^2)]^{1.5}}.
\end{equation}

Figure \ref{fig:retained} shows the location in the $f_{\rm{GW}}-M_c$ plane of all compact-object binaries found in our GC models at late times, with colors described in the figure caption. %Black systems mark BH--BH binaries, yellow BH--NS binaries, orange BH--WD binaries, red NS--NS binaries, purple NS--WD binaries, and blue WD--WD binaries. The background gray region marks the \textit{LISA} sensitivity range ($10^{-5} \, \rm{Hz} < f_{\rm{GW}} < 1 \, \rm{Hz}$).
The three solid black curves mark the boundary for detecting systems with $S/N \geq 2$ at distances of $d=9$ kpc (bottom; the median heliocentric distance of MW GCs, as calculated from \citep{Harris1996}), $d=800$ kpc (middle; the distance to Andromeda), and $d=1.6\times10^4$ kpc (top; the distance to the Virgo cluster). All points lying above these three lines will be observable by \textit{LISA} at each respective distance.

\subsection{Eccentricities}
\label{sec:eccentricities}

Systems marked as open circles in Figure \ref{fig:retained} have $e \geq 0.99$. As mentioned in \citep{Rodriguez2017}, this set of GC models implements an approximation in \texttt{BSE} that only applies gravitational radiation (GR) for sufficiently compact binaries. Under this approximation, some of these wide but highly eccentric binaries should in fact inspiral on a timescale shorter than the cluster dynamical timescale, $t_{\rm{dyn}}$, if the GR approximation is removed. In this case, the validity of these open circle systems as \textit{LISA} sources is uncertain. However, systems marked as filled circles are lower eccentricity systems with $t_{\rm{dyn}} < t_{\rm{inspiral}}$. The evolution of these systems is dominated by frequent dynamical encounters in the cluster and therefore will be unchanged by the GR approximation.

\begin{figure}
\begin{center}
\includegraphics[width=0.9\columnwidth]{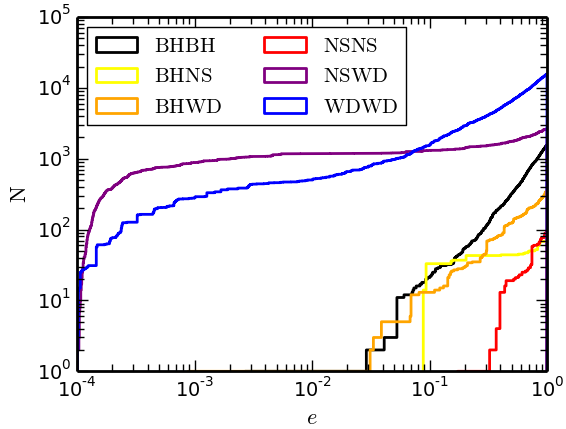}
\caption{\label{fig:ecc_retained} Cumulative distribution of eccentricity for all binaries retained at late times in GC models shown in Figure \ref{fig:retained}. The colors are listed in the figure legend and are the same as in Figure \ref{fig:retained}.}
\end{center}
\end{figure}

Figure \ref{fig:ecc_retained} shows the cumulative distribution of the eccentricities for all binaries shown in Figure \ref{fig:retained}. Many of these binaries have high eccentricities ($14\%$ of all binaries in Figure \ref{fig:retained} have $e \geq 0.1$), as expected for systems in GCs that experience frequent dynamical encounters. For binaries with $f_{\rm{GW}} \gtrsim 10^{-3}$ Hz, \textit{LISA} can measure eccentricities in excess of 0.01.

%The relatively low eccentricities exhibited in Figure \ref{fig:ecc_retained} for the NS--WD (purple curve) and WD--WD systems (blue curve) are the result of mass-transfer episodes. We adopt the treatment in \texttt{BSE} which sets $e=0$ upon onset of mass-transfer \citep{Hurley2002}.

\begin{table}
\begin{tabular}{c|c|c|c|c}
\hline
\hline
Type & All binaries & \textit{LISA} range & $S/N \geq 2$ & $S/N \geq 7$ \\
\hline
WDWD & $1.73\times 10^4$ & $1.27 \times 10^4$ & 29.3 (19.6) & 5.6 (4.3)\\
NSNS & 36.8 & 21.9 & 2.0 (1.4) & 1.2 (0.8) \\
BHBH & 215 & 33.5 & 10.2 (6.9) & 6.9 (3.9)\\
WDNS & 994 & 708 & 9.7 (5.7) & 5.5 (2.7) \\
WDBH & 84.1 & 19.4 & 3.6 (3.6) & 2.2 (2.2) \\
BHNS & 3.5 & 0 & 0 (0) & 0 (0)\\
\hline
Total & $1.87 \times 10^4$ & $1.35 \times 10^4$ & 54.8 (37.2) & 21.3 (13.8) \\
\hline
\hline
\end{tabular}
\caption{\label{table:Retained}Predicted values of different compact-object binaries in MW GCs at present day. Column 2 shows all binaries, column 3 shows number of binaries whose peak gravitational-wave frequency falls within the \textit{LISA} range ($10^{-5}\, \rm{Hz}\leq \it{f}_{\rm{GW}} \leq \rm{1} \, \rm{Hz}$), columns 4 and 5 show the number of binaries that have $S/N \geq 2$ and $S/N \geq 7$, respectively. The numbers in parentheses in columns 4 and 5 show the values if all systems with $e \geq 0.99$ are excluded, as discussed in Section \ref{sec:eccentricities}.}
\end{table}

\subsection{Estimating the total number of sources}

%Note that Figure \ref{fig:retained} shows results for \textit{all snapshots} in the range $11 \,\rm{Gyr} \leq \it{t} \leq 12 \,\rm{Gyr}$ . Because \textit{LISA} will effectively observe these clusters at only a single snapshot in time, this figure overcounts the number of binaries by a factor equal to the number of snapshots ($\sim 10$) in the specified time window (11--12 Gyr).

In order to estimate the true number of sources \textit{LISA} will observe, we rescale these numbers based upon the weighting scheme discussed in Section \ref{sec:method} to match the observed MW clusters. %Additionally, a heliocentric distance for each GC model is randomly drawn from the observed heliocentric distances of MW clusters \citep{Harris1996}, and the $S/N$ for all binaries in each model is calculated at the randomly selected distance.  We draw 10 random heliocentric distances for each model, and then re-scale our final results down by a factor of 10. Assigning distances in this manner gives us a distance distribution representative of the MW GC position distribution (as opposed to using the heliocentric distances specified for each GC model, which do not span the full MW distribution) and is permissible because the presence of \textit{LISA} sources in models at late times is independent of the heliocentric distance assumed for the model.
Table \ref{table:Retained} shows the results after this rescaling. We predict $\sim 21$ total \textit{LISA} sources to be found in the MW with $S/N \geq 7$  The two dominant populations of these will be WD--WD binaries ($\sim 6$ predicted) and BH--BH binaries ($\sim 7$ predicted). The numbers in parentheses in columns 4 and 5 of Table \ref{table:Retained} show the values if we exclude all systems with $e \geq 0.99$ (open circle systems in Figure \ref{fig:retained}) which may result from the GR approximation, as discussed in Section \ref{sec:eccentricities}. Removing these systems, we predict $\sim 14$ \textit{LISA} sources, including $\sim 4$ BH--BH binaries. We conclude that removing the GR approximation utilized in \texttt{BSE} within these models is unlikely to significantly affect our predicted numbers. Developing a full grid of models that match the MW clusters and include the updated GR approximation as well as the post-Newtonian terms described in \citep{Rodriguez2017} will be the subject of a later analysis.

\begin{figure}
\begin{center}
\includegraphics[width=0.85\columnwidth]{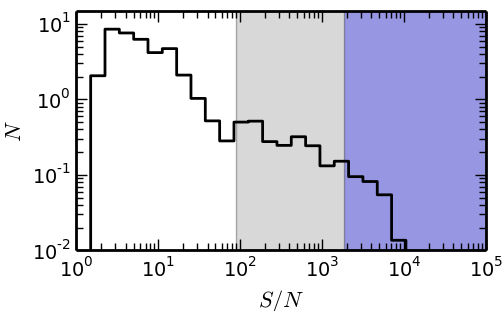}
\caption{\label{fig:SNR} Distribution of $S/N$ for all retained compact-object binaries in the MW at present, after applying the weighting scheme discussed in the text. The shaded gray (blue) region shows the MW systems with $S/N$ high enough to be observed in Andromeda (Virgo cluster).}
\end{center}
\end{figure}

%\begin{figure*}
%\begin{center}
%\includegraphics[width=0.85\textwidth]{Escaped}
%\caption{\label{fig:escaped} Same as Figure \ref{fig:retained}, but for all compact object binaries ejected from GCs, evolved forward in time to the present day as described in Section \ref{sec:models}.}
%\end{center}
%\end{figure*}

Figure \ref{fig:SNR} shows the distribution of $S/N$ for all retained compact-object binaries predicted to be observed in the MW at present.  As shown in the figure, the majority of these binaries will be observed with $S/N \lesssim 10$; however, a small number of the BH--BH binaries may be observed with $S/N$ high enough to be detected out to Andromeda or even the Virgo cluster. The shaded gray and blue regions of Figure \ref{fig:SNR} show the MW systems which would have $S/N \geq 2$ in Andromeda and the Virgo cluster, respectively, assuming a heliocentric distance of 9 kpc (the median distance of all MW GCs) for all MW sources.

Andromeda has been shown to contain approximately 500 GCs \citep[e.g.,][]{Galleti2004} and at least 12,000 GCs exist in the Virgo cluster \citep[e.g.,][]{Peng2008}. Scaling up our results to match these GC numbers, we predict, as a back-of-the-envelope estimate, that $\sim 8$ and $\sim 80$ BH--BH binaries will be resolvable by \textit{LISA} with $S/N \geq 2$ in Andromeda and the Virgo cluster, respectively.

In addition to compact-object binaries \textit{retained} in their host clusters at late times, we also considered all compact object-binaries \textit{ejected} from our GC models.  We predict that ejected systems will contribute only a few MW sources resolvable with $S/N \geq 2$. It is unlikely any of these systems will have sufficiently high $S/N$ to be detected in Andromeda or the Virgo cluster. More details on the ejected binaries can be found in the Appendix.

\section{Conclusion}
\label{sec:conclusion}
We have explored the formation of double-compact-object binaries in GCs that may be detectable by \textit{LISA}. %Using a set of 137 fully-evolved globular cluster models that effectively match the properties of observed globular clusters in the MW.
We predict $\sim 55$ total sources will be detectable in MW GCs with $S/N \geq 2$ and  $\sim 21$ systems with $S/N \geq 7$ by \textit{LISA}. Furthermore, we predict tens of additional sources will be resolvable within Andromeda and the Virgo cluster.

%Additionally, we showed that ejected systems do not contribute as significantly to the \textit{LISA} population as the retained binaries.%, but we still predict a few of these ejected systems will be resolvable by \textit{LISA}.

Because the majority of these dynamically formed systems are retained within their host GCs, localization to a particular GC may be possible for sources with sufficiently high $S/N$. Furthermore, the moderate to high eccentricity of these dynamically formed systems will likely make them distinguishable from \textit{LISA} sources formed in the Galactic field. 

In this analysis we have explored only the contribution of the $\sim 150$ \textit{observed} GCs  in the MW. However, the MW may contain as many as $\sim 200$ total GCs, with  $\sim 1/4$ of them lying deep in the Galactic plane making them difficult to observe. The contribution of these unobserved clusters would boost our predicted number of \textit{LISA} sources in the MW by $\sim 25\%$.

Furthermore, the many more less massive clusters that would have formed together with the present-day surviving clusters are likely to contribute to the total number of sources since they also leave binaries that would now populate the Galactic field \citep[e.g.,][]{Brockamp2014}. Investigation of dissolved is outside the scope of this Letter. The number of \textit{LISA} sources produced in \textit{present day} GCs could be viewed as a lower limit to the total number of dynamically formed sources in the Galaxy.

In future work, we intend to perform a side-by-side comparison of the dynamical formation channels explored in this analysis and field formation channels of \textit{LISA} binaries to predict the total number of \textit{LISA} sources in the MW. Additionally, we intend to explore the contribution of post-Newtonian effects to the formation and evolution of \textit{LISA} binaries in GCs.

%%%%%%%%%%%%%%%%%%%%%%%%%%%%%%%%%

\begin{acknowledgments}
This work was supported by NASA ATP Grant NNX14AP92G 
and NSF Grant AST-1716762. K.K. acknowledges support by the National Science Foundation Graduate Research Fellowship Program under Grant No. DGE-1324585.
S.C. acknowledges support from
CIERA, the National Aeronautics and Space Administration
through a Chandra Award Number TM5-16004X/NAS8-
03060 issued by the Chandra X-ray Observatory Center
(operated by the Smithsonian Astrophysical Observatory for and on behalf of the National Aeronautics
Space Administration under contract NAS8-03060), 
and Hubble Space Telescope Archival research 
grant HST-AR-14555.001-A (from the Space Telescope 
Science Institute, which is operated by the Association of Universities for Research in Astronomy, Incorporated, under NASA contract NAS5-26555).
\end{acknowledgments}

%%%%%%%%%%%%%%%%%%%%%%%%%%%%%%
\appendix

\section{Calculation of the signal-to-noise ratio}
\label{sec:SNR}

To compute the \textit{LISA} sensitivity for the binaries considered in this analysis, we model GWs at the leading quadrupole order. The (angle-averaged) signal-to-noise ratio ($S/N$) at which a given binary can be detected is given by

\begin{equation}
(S/N)^2 = \sum_{n} \langle (S/N)_n^2 \rangle = \sum_n \int \Big[ \frac{h_{n}(f_n)}{h_{f}(f_n)} \Big]^2 d \ln{f_n}.
\end{equation}
Here $n$ labels the harmonics at frequency $f_n \simeq n\,f_{\rm{orb}}$, where $f_{\rm{orb}}$ is the orbital frequency. $h_f$ is the spectral amplitude value for a specified gravitational-wave frequency. Assuming, as a basic approximation, that binaries do not chirp appreciably over the course of a single \textit{LISA} observation, we can approximate $S/N$ as:
\begin{equation}
(S/N)^2 = \sum_{n=1}^{\infty} \langle (S/N)_n^2 \rangle \approx \sum_{n=1}^{\infty} \frac{h_{n,\rm{o}}^2}{h_{f}^2} \frac{g(n,e)}{n^2} T_{\rm{obs}}.
\end{equation}
Here $g(n,e)$ is given by equation (20) in \citep{Peters1963}, $T_{\rm{obs}}= 4$ years is the \textit{LISA} observation time, and $h_{n,\rm{o}}$ is given by

\begin{equation}
\label{eq:h_o}
h_{n,\rm{o}} = \frac{G}{c^2}\frac{M_c}{D}\left(\frac{G}{c^3}\pi \,f_n \, M_c \right)^{2/3},
\end{equation}
where $D$ is the distance of the binary from the detector and $M_c$ is the chirp mass,

\begin{equation}
\label{chirp}
M_c = \frac{(M_1 M_2)^{3/5}}{(M_1 + M_2)^{1/5}},
\end{equation}
where $M_1$ and $M_2$ are the component masses of the binary.

\section{Ejected binaries}

\begin{table}
\begin{tabular}{c|c|c|c|c}
\hline
\hline
Type & All binaries & \textit{LISA} range & $S/N \geq 2$ & $S/N \geq 7$ \\
\hline
WDWD & $436$  & 364 & 0.4 (0.4) & 0 (0)\\
NSNS & 18.9 & 16.8 & 0.5 (0.5) & 0.3 (0.3) \\
BHBH & $4.38 \times 10^3$ & 544 & 0.9 (0.9) & 0.2 (0.2) \\
WDNS & 90.1 & 39.3 & 0.2 (0.2) & 0 (0) \\
WDBH & 13.7 & 0 & 0 (0) &0 (0)\\
BHNS & 15.3 & 4.6 & 0 (0) &0 (0) \\
\hline
Total & $4.96 \times 10^3$ & 969 & 1.9 (1.9) & 0.5 (0.5) \\
\hline
\hline
\end{tabular}
\caption{\label{table:Escaped} Same as Table \ref{table:Retained}, but for all ejected compact object binaries.}
\end{table}

\begin{figure*}
\begin{center}
\includegraphics[width=0.9\textwidth]{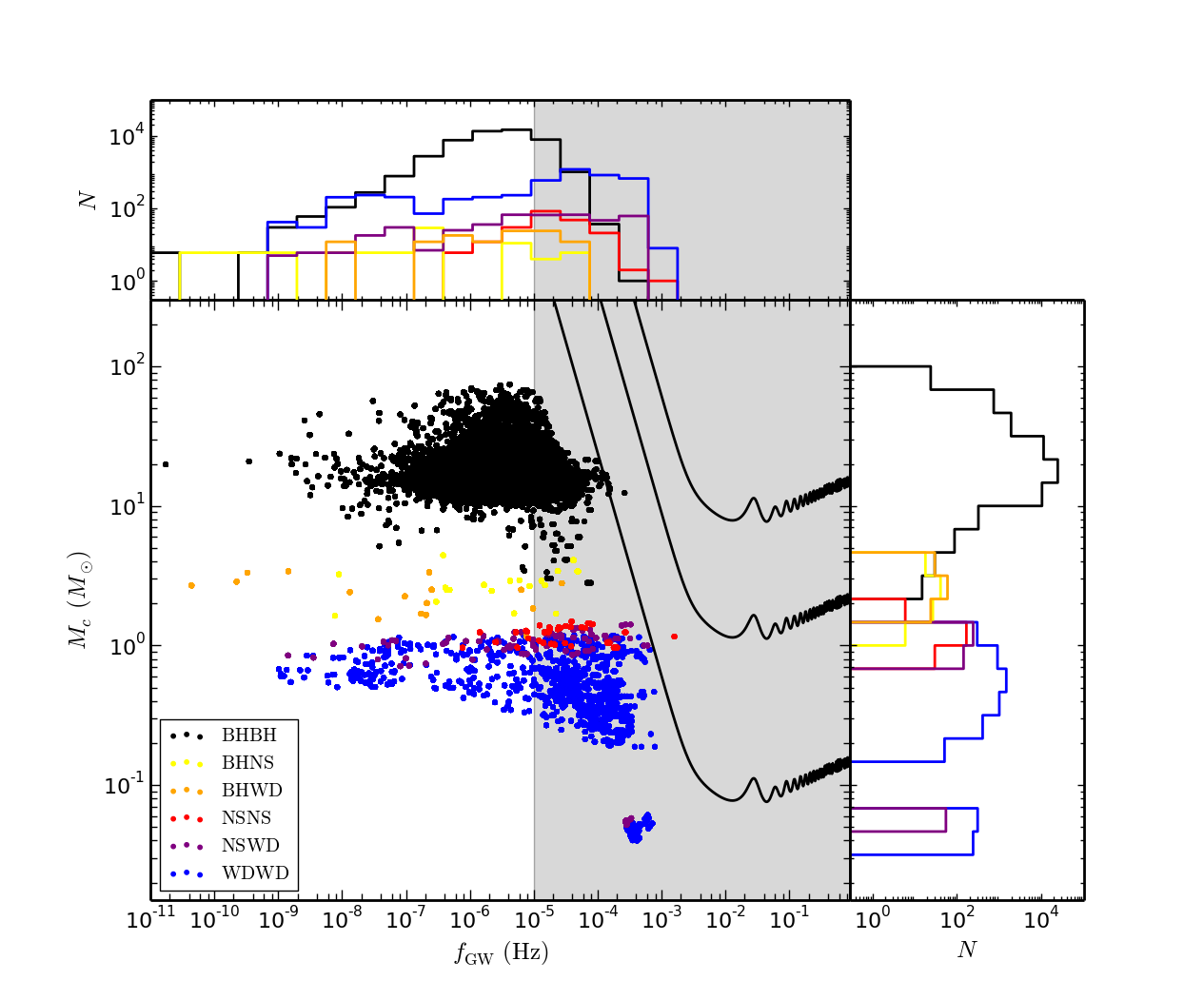}
\caption{\label{fig:escaped} Same as Figure \ref{fig:retained}, but for all compact object binaries ejected from GCs, evolved forward in time to the present day as described in Section \ref{sec:method}.}
\end{center}
\end{figure*}

Figure \ref{fig:escaped} shows the location in the $f_{\rm{GW}}-M_c$ plane of all compact object binaries ejected from our GC models and evolved forward in time to the present day using the method described in Section \ref{sec:method}. All colors here are as in Figure \ref{fig:retained}.

Comparison of Figures \ref{fig:escaped} and \ref{fig:retained} shows that ejected BH--BH binaries exhibit a much larger range in $M_c$ relative to the retained BH--BH binaries. This is consistent with our understanding of the evolution of BH binaries in GCs. Upon formation, the most massive BHs will rapidly sink to their host cluster's central region, where they undergo frequent dynamical encounters leading to binary formation, hardening, and, ultimately, ejection from the cluster. After the most massive BHs are ejected, the next most massive BHs will sink to the center, and ultimately get ejected themselves. In this manner, the average mass of BHs in a cluster is expected to decrease with the cluster's age, so that only lower mass BHs remain at late times. This explains the relatively small range in $M_c$ for the retained BH--BH binaries shown in Figure \ref{fig:retained}. But the population of ejected binaries which have not yet merged by the present day, and are therefore potential \textit{LISA} sources, will include the massive BH binaries ejected in the first few Myr of the evolution of their host clusters. If these massive BH--BH binaries are ejected with sufficiently high orbital separations, they will not inspiral by the present-day and will contribute to the relatively wide range in $M_c$ we see for the BH--BH binaries in Figure \ref{fig:escaped}.

Table \ref{table:Escaped}, which is analogous to Table \ref{table:Retained}, shows the predicted numbers of compact object binaries of each type which have been ejected by the MW GC system. Unlike systems retained within their host clusters, we predict that ejected systems will contribute only a few MW sources resolvable with $S/N \geq 2$ and it is unlikely any of these systems will be detected in Andromeda or the Virgo cluster.

It is not surprising that the numbers predicted for retained systems is larger than for ejected systems. Once a binary is ejected, GR is the only mechanism to drive the system to higher frequencies. But for systems retained within their host clusters, frequent dynamical encounters (particularly for the BH binaries which are most likely to be found in the dense cluster core, where dynamical encounters are most frequent) will frequently alter the populations of these systems, creating new systems over time and providing a mechanism to harden these binaries in addition to hardening caused by GR.

\end{document}